\documentclass{Interspeech}

\interspeechcameraready

\title{LLM-based Generative Error Correction for Rare Words with Synthetic Data and Phonetic Context}

\author[affiliation={1}]{Natsuo}{Yamashita}
\author[affiliation={1}]{Masaaki}{Yamamoto}
\author[affiliation={1}]{Hiroaki}{Kokubo}
\author[affiliation={1}]{Yohei}{Kawaguchi}

\affiliation{}{Hitachi, Litd.}{Japan}
\email{\{natsuo.yamashita.gh, maaaki.yamamoto.af, hiroaki.kokubo.dz, yohei.kawaguchi.xk\}@hitachi.com} 

\usepackage{comment}
\usepackage{arydshln}
\usepackage{booktabs} 
\usepackage{multirow} 
\usepackage{graphicx} 
\usepackage{tipa}
\usepackage{amsmath}
\usepackage{physics}
\newcommand{\vect}[1]{\mbox{\boldmath $#1$}}
\begin{document}

\maketitle

\begin{abstract}

Generative error correction (GER) with large language models (LLMs) has emerged as an effective post-processing approach to improve automatic speech recognition (ASR) performance. However, it often struggles with rare or domain-specific words due to limited training data. Furthermore, existing LLM-based GER approaches primarily rely on textual information, neglecting phonetic cues, which leads to over-correction. To address these issues, we propose a novel LLM-based GER approach that targets rare words and incorporates phonetic information. First, we generate synthetic data to contain rare words for fine-tuning the GER model. Second, we integrate ASR's N-best hypotheses along with phonetic context to mitigate over-correction. Experimental results show that our method not only improves the correction of rare words but also reduces the WER and CER across both English and Japanese datasets.
\end{abstract}

\begin{figure*}[t]
\begin{minipage}[b]{1.0\linewidth}
  \centering
  \centerline{\includegraphics[width=1.0\columnwidth]{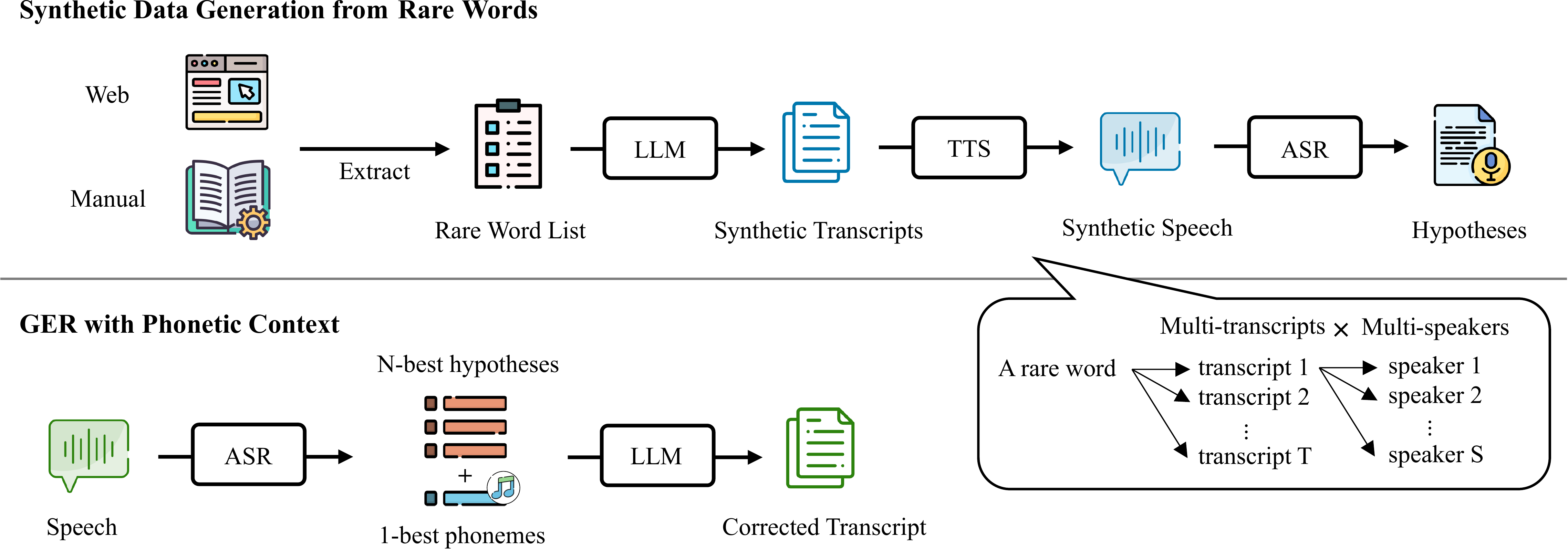}}
\end{minipage}
\caption{Overview of the proposed methods of synthetic data generation from rare words and GER with phonetic context.}
\label{fig:proposal}
\end{figure*}

\section{Introduction} \label{sec:introduction}
Automatic speech recognition (ASR) technology has achieved remarkable progress over the past decades, driven by advancements in deep learning techniques \cite{end_to_end}. 
However, these systems often produce transcription errors, particularly due to background noise, speaker accents, different speaker styles, and domain-specific terms. 
To address this issue, researchers have explored various generative error correction (GER) methods, utilizing pre-trained language models (LMs) \cite{phone_conditioned,lm_1,lm_2} and large language models (LLMs) \cite{llm_1}. 
These approaches aim to enhance ASR transcription quality by correcting grammatical, syntactical, or semantic errors. 
Recent studies \cite{multi_stage,llm_ger_1,llm_ger_2,llm_ger_3,llm_ger_4,llm_ger_5} have demonstrated the effectiveness of LLMs in processing N-best hypotheses from ASR outputs to generate refined transcriptions. 
While these methods have shown promising results, they struggle with handling rare words that are absent from their training data \cite{dictionary,llm_rare_words}. 
This limitation arises because existing GER models often rely on pre-trained datasets that lack sufficient examples of specialized terminology or rare word errors, and obtaining domain-specific error data for fine-tuning remains a challenge.

Another critical limitation of existing methods is their reliance solely on textual information while ignoring phonetic cues \cite{acoustic_llm}. 
This often results in over-correction, where unnecessary changes are made, shifting the transcription toward formal written language and reducing the fidelity of the spoken input \cite{multi_stage}. 
For instance, models may produce semantically plausible outputs that fail to preserve the original pronunciation or intent, leading to phonetically incorrect results and compromising the overall quality of ASR systems.

To overcome these challenges, this study investigates an LLM-based GER approach that specifically targets rare words and incorporates phonetic information.
First, we propose generating transcripts from rare words using an LLM and synthesizing speech from them to prepare error-pair data.
By generating multiple transcripts and synthesizing speech with multiple speakers, we ensure a variety of error patterns in the generated data, which are then used to fine-tune the LLM. 
This approach enables the model to learn from various errors, allowing it to handle both common and rare words that are not part of its initial training data. 
In contrast to previous studies \cite{multi_stage,llm_ger_4,llm_ger_5,llm_ger_6} that replace words in hypotheses or rescore hypotheses to limit the introduction of new words and avoid over-correction, our method focuses on the accurate correction of errors, even when they are not present in the hypotheses.

Second, inspired by the previous work on phone-conditioned LM  \cite{phone_conditioned}, we incorporate phonetic information into the LLM-based GER process to improve both phonetic and semantic correctness.
Unlike the previous work \cite{acoustic_llm}, which integrated acoustic features into the ASR decoding process, our approach focuses on utilizing phonetic context derived from ASR hypotheses.
This approach is significantly easier to integrate with the latest LLMs and ASR models that are continuously updated, especially given that the internal structures of some models, such as ChatGPT \cite{chatgpt}, are not publicly disclosed.
By leveraging an LLM to predict simplified phonetic readings of text, we provide additional phonetic context as input to our correction model. This enables the system to incorporate both textual and phonetic features, resulting in more accurate corrections that better align with the original spoken input.

Experimental results demonstrate that our approach 
improves the correction of rare words, and reduces the word error rate (WER) and character error rate (CER) across both English and Japanese datasets.


\section{Related Work}
\subsection{Error correction for rare words}
In many practical scenarios, a rare word list can be obtained in advance from sources such as user names, meeting chat logs, websites, manuals, or even words registered by users \cite{inter_biasing}.
Several studies \cite{ed-cec,entity_resolution} have explored the use of rare word lists for error correction, but the error-pair data is limited within the training data.
A recent study proposed an RAG-based keyword error correction using LLMs by searching for similar entries in rare word lists \cite{retrieval}, but correcting non-keyword errors remains a challenge.
In this paper, we propose generating diverse error-pair data that includes rare words, enabling the error correction of both rare and non-rare word errors.

\subsection{Phonetic notation}

In the field of linguistics, the International Phonetic Alphabet (IPA) \cite{ipa} has been widely used to accurately represent the pronunciation of all languages around the world. However, it is often perceived as complex, particularly by non-native speakers, due to the specialized linguistic symbols \cite{simple_ipa}. 
In the field of text-to-speech (TTS), to represent phonetic information, ARPAbet for English and romanized Kana for Japanese are commonly used by using graphene-to-phoneme (G2P) tools \cite{phone_conditioned,tacotron2,g2p_tts}.
A very recent study investigated LLM-based grapheme-to-phoneme conversion and reported that the performance of an LLM alone is significantly lower than that of an approach combining an LLM and a G2P dictionary \cite{llm-g2p}. 
This finding suggests that current LLMs struggle to effectively handle specialized phonetic representations.
To address this issue, we propose an LLM-based approach using simplified phonetic representation to enable LLMs to better comprehend phonetic information.




\section{Methodology}

Figure \ref{fig:proposal} illustrates our proposed methods. 
First, we propose training a GER model using synthetic data generated by gerative models based on a given rare word list. Additionally, to prevent over-correction caused by ignoring phonetic information, we propose explicitly providing the GER model not only N-best hypotheses, but also 1-best phonetic representations.

\subsection{Synthetic data generation from rare words}

Given a set of $W$ biasning words $\left\{\vect{w_n}\right\}_{n=1}^{W}$, we generate $T$ synthesized transcripts $\left\{\vect{t}^{(w)}_{n}\right\}_{n=1}^{T}$ that include each rare word. To create these transcripts, we prompt an LLM with instructions, optionally providing contextual information when available. For instance, the instruction might read: \textit{``Provide 5 different English sentences in various contexts that include the term $\vect{w_n}$, which is a medical term.''}.
Subsequently, using text-to-speech models, we generate $S$ synthetic spoken utterances $\left\{\vect{u}^{(w, t)}_{n}\right\}_{n=1}^{S}$ for each synthetic transcript.
Here, multiple transcripts are generated for each rare word to capture a variety of contexts, and utterances are synthesized with multiple speakers to ensure diversity in speech styles. 
Finally, using an ASR model, we obtain ${W}\times{T}\times{S}$ hypotheses paired with the corresponding transcripts to fine-tune a GER model.
Note that hypotheses containing no errors are excluded, while those with errors only in non-rare words are included in the training process to reduce the training cost.

\begin{table}[t]
    \centering
    \caption{Examples of English phonetic representations.}
    \label{tbl:phoneme}
    \resizebox{0.90\linewidth}{!}{
        \tabcolsep = 6pt
        \begin{tabular}{ll}
            \toprule
            Ground truth & the sun is rising \\
            Hypothesis & the son is rising \\
            \midrule
            IPA & \textipa{ð@ s@n Iz "raIzIN}   \\
            ARPAbet & DHAH0 SAH1N IH1Z RAY1ZIH0NG \\
            LSP & thuh sun iz rahy-zing \\
            \bottomrule
        \end{tabular}}
\end{table}

\subsection{GER with phonetic context}
As discussed in Section \ref{sec:introduction}, GER models often suffer from over-correction toward semantic correctness, ignoring phonetic information. 
To address this challenge, we propose combining the N-best hypotheses with the phonetic representations derived from the 1-best hypothesis of the ASR model. 
For phonetic representations, we investigate the use of IPA which is an international standard for representing pronunciation, as well as ARPAbet and romanized Kana, which are commonly used in the TTS field.
Additionally, for LLM-based GER, we introduce a novel phonetic representation called LLM-based Simplified Phoneme (LSP).
We hypothesize that a simpler phonetic representation is sufficient to capture phonetic information, as the aforementioned common representations include many specialized expressions that are difficult to intuitively understand. 
Furthermore, an LLM is more likely to effectively align transcripts with LSP because it outputs phonetic representations that it has learned.
The prompting templates for LSP generation are as follows: \textit{``Convert the English text to simplified pronunciation.''} or \textit{``Convert the Japanese text to simplified Kana-like pronunciation.''}.
As Table \ref{tbl:phoneme} shows, LSP is easier to understand compared to IPA and ARPAbet.
Subsequently, the 1-best phonemes are combined with N-best hypotheses from the ASR model to train the LLM. 
Here, we focus on utilizing 1-best phonemes rather than N-best to minimize computational cost and avoid complex input to LLMs.



\section{Experimental Setup}


\begin{table}[t]
    \centering
    \caption{Summary of evaluation datasets.}
    \label{tbl:datasets}
    \resizebox{1.0\linewidth}{!}{
        \tabcolsep = 3pt
        \begin{tabular}{lcccccc}
            \toprule
            \multirow{2}{*}{dataset}&
            \multirow{2}{*}{lang.}&
            \multirow{2}{*}{\# utts.}&rare words&TTS&TTS\\
            & &  & coverage (\%) &  model & train-data\\
            \midrule
            LibriSpeech & EN & 2620 & 2.3 & - & - \\
            EDGAR & EN & 500 & 5.8 & VITS & VCTK \\
            \addlinespace[0.05cm]
            \hdashline
            \addlinespace[0.1cm]
            CSJ eval1 & JA & 1411 & 5.0 & - & - \\
            CSJ eval2 & JA & 1420 & 4.5 & - & - \\
            MedTxt & JA & 605 & 10.0 & FastSpeech2 & JSUT \\
            \bottomrule
        \end{tabular}}
\end{table}

\subsection{Dataset}


To investigate the efficiency of our approach in a multilingual setting, we evaluated both English and Japanese datasets, as summarized in Table \ref{tbl:datasets}.
Since public datasets like Common Voice \cite{common_voice} typically contain very few rare words and are often included in the training sets of foundation models, we constructed synthetic datasets that focus on rare words for evaluation alongside public datasets.
\begin{itemize}
    \item \textbf{LibriSpeech} \cite{librispeech}: A dataset of 960 hours of English audiobook recordings. We used the \textit{test-clean} for evaluation.
    \item \textbf{EDGAR} \cite{edgar}: An English text corpus of annual reports from SEC EDGAR filings (1993–2020). We extracted 500 sentences from it and synthesized speech using VITS \cite{vits}, which was trained on the VCTK corpus \cite{vctk}.
    \item \textbf{Corpus of Spontaneous Japanese (CSJ)}: A collection of Japanese public speech on academic topics. We evaluated \textit{eval1} and \textit{eval2}, which contain a sufficient number of rare words, prepared using the Kaldi recipe \cite{kaldi}.
    \item \textbf{MedTxt} \cite{medtxt}: A Japanese text corpus of case reports extracted via OCR from J-Stage open-access PDF articles. We partially used words tagged as \textit{disease}, \textit{anatomical part}, or \textit{feature} as rare words. Synthetic speech was generated using FastSpeech2 \cite{fastspeech2} and HiFi-GAN \cite{hifi_gan}, which were trained on the JSUT corpus \cite{jsut}.
\end{itemize}



\subsection{Constructing rare word lists}
Due to the lack of available rare word lists for the LibriSpeech, EDGAR, and CSJ datasets, we constructed rare word lists for each dataset with the assistance of an LLM prompt, such as \textit{``Extract highly complex words for recognition, including technical terms, names of people, and names of places.''}. 
To ensure fair experimental conditions, we prepared the rare word lists in accordance with the experimental setup of previous studies \cite{ed-cec,contextualized}, ensuring that the coverage of rare words remains below 10\% of the total words used during inference, as shown in Table \ref{tbl:datasets}. 
To promote reproducibility and support future research, we have publicly released the rare word lists\footnote{\scriptsize{https://github.com/natsuooo/llm-ger}}.

\subsection{Generating synthetic data}
Due to the limitations of our computing resources, we prepared data with $T=4$ and $S=7$ for the experiments, which was split into training and validation datasets at a ratio of 4:1.
Given the absence of open-source, high-quality Japanese TTS models, we utilized Microsoft Azure Text-to-Speech \cite{azure} for TTS.
For the ASR model, we employed a popular end-to-end ASR system, Whisper (Whisper-Large-v3-turbo) \cite{whisper}, to prepare simulated data and conduct inference. No domain-adaptation techniques were applied.
We obtained N-best hypotheses from the beam search process of Whisper, setting $N=5$.

\subsection{Generating phonetic representations}
We used the Python tool eng\_to\_ipa\footnote{\scriptsize{\url{https://pypi.org/project/eng-to-ipa}}}  to convert English text to IPA. Due to the absence of a common tool for converting Japanese text to IPA, we employed an LLM to perform the conversion based on the existing work \cite{llm-g2p}. For TTS-phoneme representations, we used the G2P tool\footnote{\scriptsize{\url{https://github.com/Kyubyong/g2p}}} for ARPAbet in English and pyopenjtalk\footnote{\scriptsize{\url{https://github.com/r9y9/pyopenjtalk}}}
for romanized Kana in Japanese. 

\subsection{Training LLM-based GER models}
We employed ChatGPT-4o-mini (2024-07-18) \cite{chatgpt} for both the English and Japanese datasets, Llama-3.3-70B-Instruct\footnote{\scriptsize{\url{https://huggingface.co/meta-llama/Llama-3.3-70B-Instruct}}} for the English datasets, and Llama-3.1-Swallow-70B-Instruct-v0.3\footnote{\scriptsize{\url{https://huggingface.co/tokyotech-llm/Llama-3.1-Swallow-70B-Instruct-v0.3}}} for the Japanese datasets.
We fine-tuned ChatGPT with 3 epochs, a batch size of 10, and a learning rate multiplier of 1.8. 
For fine-tuning the Llama models, we employed 8 NVIDIA H100 GPUs with 10 epochs, a batch size of 4, a learning rate of 5.0e-5, and the Adam optimizer. To improve computational efficiency, we adopted the low-rank adaptation (LoRA) method \cite{lora} with a rank of 16, applied to all linear modules.

\subsection{Evaluation metrics}
We evaluated the results with WER for the English datasets and CER for the Japanese datasets. 
Additionally, we reported the recall and the precision of the rare words, following the methodology described in \cite{spell_my_name}. 
A high recall indicates that the model successfully identifies most of the rare words, while a high precision means that when the model identifies a word as a rare word, it is highly likely to be correct.
Given that the proportion of rare words is very small in the overall text, our primary objective is to first improve WER and CER. Subsequently, we aim to enhance recall without causing significant degradation in precision.



\section{Results and Analysis}

\begin{table*}[t]
    \centering
    \caption{Results with \textbf{ChatGPT}. Reported metrics are in the following formats: ``WER / recall / precision'' for the English datasets (LibriSpeech and EDGAR) and  ``CER / recall / precision'' for the Japanese datasets (CSJ and MedTxt).}
    \label{tbl:chatgpt}
    \resizebox{0.92\linewidth}{!}{
    \tabcolsep = 5pt
    \begin{tabular}{@{}lccccccc@{}}
    \toprule
    \addlinespace[0.1cm]
    \addlinespace[0.1cm]
    method                       &     LibriSpeech     &        EDGAR        &         CSJ eval1    &       CSJ eval2      &         MedTxt      \\
    \midrule
    Whisper-large-v3-turbo       &2.7\,/\,80.5\,/\,\textbf{98.7}&16.0\,/\,74.7\,/\,92.5&15.5\,/\,44.5\,/\,98.8&14.8\,/\,55.6\,/\,99.6&18.8\,/\,27.6\,/\,99.1 \\
    Baseline  1:  prompt only    &2.7\,/\,81.6\,/\,\textbf{98.7}&14.7\,/\,77.5\,/\,93.7&15.4\,/\,48.3\,/\,98.9&14.8\,/\,56.0\,/\,99.3&17.6\,/\,29.8\,/\,\textbf{99.2} \\
    Baseline  2:  N-best         &2.8\,/\,81.1\,/\,98.6&15.1\,/\,76.1\,/\,93.6&15.7\,/\,49.6\,/\,98.6&14.7\,/\,58.8\,/\,\textbf{99.7}&17.6\,/\,30.5\,/\,\textbf{99.2} \\
    \midrule
    Synth.\,+\,N-best            &2.6\,/\,91.7\,/\,97.7&14.1\,/\,81.0\,/\,\textbf{95.6}&14.2\,/\,81.1\,/\,98.6&\textbf{12.6}\,/\,81.7\,/\,98.3&8.6\,/\,85.0\,/\,97.8 \\
    Synth.\,+\,N-best\,+\,IPA    &2.8\,/\,92.4\,/\,96.6&13.9\,/\,79.5\,/\,95.0&27.3\,/\,\textbf{90.5}\,/\,91.9&22.3\,/\,83.3\,/\,80.1&9.4\,/\,84.7\,/\,94.9 \\
    Synth.\,+\,N-best\,+\,TTS-phoneme&2.7\,/\,91.3\,/\,97.3&14.0\,/\,80.9\,/\,95.1&13.8\,/\,78.1\,/\,\textbf{99.1}&13.6\,/\,79.8\,/\,98.0&\textbf{7.8}\,/\,\textbf{88.6}\,/\,94.3 \\
    Synth.\,+\,N-best\,+\,LSP    &\textbf{2.5}\,/\,\textbf{94.2}\,/\,96.4&\textbf{13.8}\,/\,\textbf{81.1}\,/\,95.1&\textbf{13.7}\,/\,82.0\,/\,98.9&\textbf{12.6}\,/\,\textbf{84.9}\,/\,98.3&\textbf{7.8}\,/\,87.4\,/\,97.0  \\
    \bottomrule
    \end{tabular}
    }
\end{table*}

\begin{table*}[t]
    \centering
    \caption{Results with the \textbf{Llama models}. Reported metrics are in the following formats: ``WER / recall / precision'' for the English datasets (LibriSpeech and EDGAR) and  ``CER / recall / precision'' for the Japanese datasets (CSJ and MedTxt).}
    \label{tbl:llama}
    \resizebox{0.92\linewidth}{!}{
    \tabcolsep = 5pt
    \begin{tabular}{@{}lccccccc@{}}
    \toprule
    \addlinespace[0.1cm]
    \addlinespace[0.1cm]
    method                       &      LibriSpeech    &         EDGAR        &        CSJ eval1     &        CSJ eval2     &        MedTxt         \\
    \midrule
    Whisper-large-v3-turbo       &2.7\,/\,79.9\,/\,97.4&16.0\,/\,74.7\,/\,92.5&15.5\,/\,44.5\,/\,98.8&14.8\,/\,55.6\,/\,99.6&18.8\,/\,27.6\,/\,99.1 \\
    Baseline  1:  prompt only    &2.8\,/\,80.3\,/\,96.8&15.6\,/\,77.1\,/\,94.4&15.8\,/\,46.4\,/\,97.8&15.1\,/\,55.6\,/\,99.6&18.0\,/\,31.0\,/\,\textbf{99.2} \\
    Baseline  2:  N-best         &2.7\,/\,82.1\,/\,97.4&14.6\,/\,73.8\,/\,\textbf{98.1}&15.6\,/\,47.5\,/\,\textbf{98.9}&14.7\,/\,58.8\,/\,\textbf{99.7}&17.8\,/\,20.8\,/\,\textbf{99.2} \\
    \midrule
    Synth.\,+\,N-best            &2.9\,/\,\textbf{91.4}\,/\,\textbf{98.2}&\textbf{14.2}\,/\,\textbf{78.2}\,/\,94.2&14.3\,/\,62.9\,/\,98.2&14.6\,/\,65.9\,/\,\textbf{99.7}&10.3\,/\,84.7\,/\,92.8 \\
    Synth.\,+\,N-best\,+\,IPA    &3.5\,/\,90.8\,/\,95.6&14.9\,/\,77.3\,/\,94.0&25.2\,/\,\textbf{72.4}\,/\,86.5&18.1\,/\,\textbf{79.2}\,/\,87.1&21.9\,/\,76.5\,/\,83.4 \\       
    Synth.\,+\,N-best\,+\,TTS-phoneme&3.0\,/\,84.8\,/\,97.4&14.4\,/\,77.7\,/\,94.9&\textbf{14.1}\,/\,66.5\,/\,98.1&12.9\,/\,66.6\,/\,99.1&10.4\,/\,84.3\,/\,93.0 \\
    Synth.\,+\,N-best\,+\,LSP    &\textbf{2.6}\,/\,86.5\,/\,97.7&\textbf{14.2}\,/\,77.4\,/\,94.3&\textbf{14.1}\,/\,65.7\,/\,98.6&\textbf{12.8}\,/\,64.1\,/\,99.3&\textbf{9.3}\,/\,\textbf{85.0}\,/\,92.1  \\
    \bottomrule
    \end{tabular}%
    }
\end{table*}

\subsection{Results}


Table \ref{tbl:chatgpt} and \ref{tbl:llama} present the WER for the English datasets, the CER for the Japanese datasets, as well as the recall and precision scores when using ChatGPT or the Llama models as the LLMs.
First, we note that most of the results using ChatGPT were better than those using the LLama models.
The initial transcription results from Whisper reveal a very low recall for rare words, particularly in Japanese, 
which has numerous homophones of Chinese characters due to its limited variety of phonemes \cite{japanese_homonyms}.
In Baseline 1, which applies a correction method based solely on prompts and 1-best hypothesis, and Baseline 2, which utilizes N-best hypotheses for correction \cite{llm_ger_3}, the improvements in the WER, CER, and recall were modest. 
For example, in the case of the CSJ eval1 dataset using Baseline 2 with the LLama model, the CER remained almost unchanged from 15.5\% to 15.6\% and the recall was slightly improved from 44.5\% to 47.5\%.

In the proposed method, which leverages synthetic data and N-best hypotheses but excludes phonetic context, we observed significant improvements in the WER, CER, and recall, without a significant decrease in precision across most datasets.
Notably, the improvement in recall was particularly substantial for Japanese which contains many homophones.
For example, when using ChatGPT for MedTxt dataset which includes many complex medical terms, the recall dramatically increased from 27.6\% to 85.0\%.
These results clearly show that the GER models effectively learned to correct both rare and non-rare words from synthetic data generated from rare words.


Next, we discuss the impact of different phonetic representations.
When using IPA, the complexity of its representations caused the GER models to overfit to the rare words, resulting in high recall but a substantial decline in the WER and CER, particularly in Japanese datasets.
When using TTS-phonemes, such as ARPAbet for English and romanized Kana for Japanese, were used, no observable improvement was found in English datasets, while a slight improvement was observed in the Japanese datasets.
In contrast, compared to the proposed method without phonetic context, incorporating LSP resulted in the further improvement in the WER and CER across all datasets, without significantly degrading recall or precision.
Specifically, for the LibriSpeech dataset using LLama, though the WER of Whisper increased from 2.7\% to 2.9\% without phonetic context, leveraging LSP improved the WER to 2.6\%. 
These results indicate that LSP helped prevent over-correction caused by an excessive focus on semantic aspects of texts and thereby improved the WER and CER.

\subsection{Analysis of the number of transcripts and speakers}
\begin{figure}[t]
\begin{minipage}[b]{0.93\linewidth}
  \centering
  \centerline{\includegraphics[width=0.93\columnwidth]{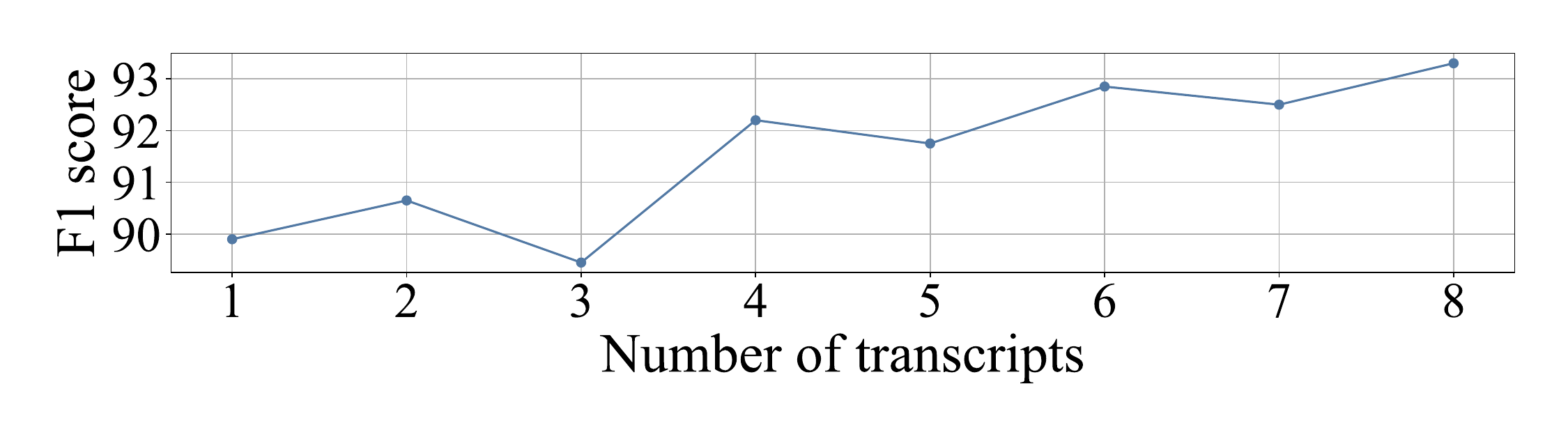}}
\end{minipage}
\begin{minipage}[b]{0.93\linewidth}
  \centering
  \centerline{\includegraphics[width=0.88\columnwidth]{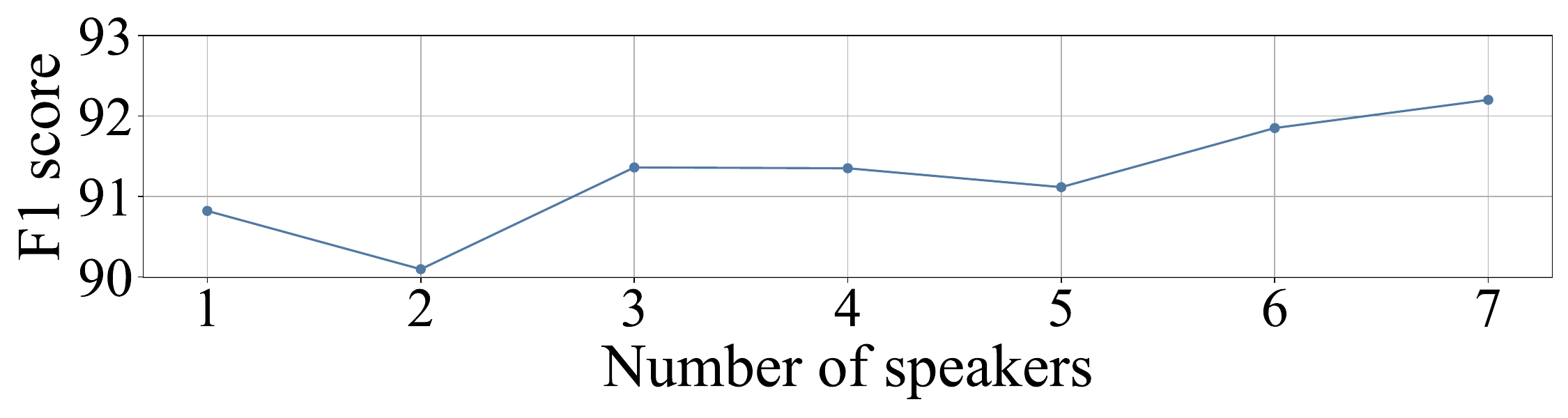}}
\end{minipage}
\caption{F1 scores of rare words with different numbers of transcripts and speakers in the Medtxt dataset using ChatGPT.}
\label{fig:number}
\end{figure}

Figure \ref{fig:number} illustrates how the number of transcripts and speakers during synthetic data generation impacts the rare word recognition performance in the MedTxt dataset when using ChatGPT.
As shown in Figure \ref{fig:number}, a significant improvement of approximately 2\% in the F1 score is observed when the number of transcripts increases from 1 to 4 while keeping the number of speakers fixed at 7. Beyond this point, the rate of improvement gradually diminishes. Regarding the number of speakers with the number of transcripts fixed at 4, although the improvement is smaller compared to increasing the number of transcripts, increasing the number of speakers and thereby increasing the variation in speaker styles contributes to enhancing the F1 scores.

\section{Conclusions}

In this paper, we proposed an approach to enhance the LLM-based GER for transcripts containing rare words. 
We introduced a method for generating diverse synthetic data containing rare words, combined with leveraging LLM-based simplified phonemes to avoid over-correction.
The experimental results demonstrated that our approach achieved the best WER and CER across all datasets, while improving the recall of rare words without compromising precision. 
Future work includes evaluating our method on larger datasets and across different domains to assess its scalability and generalizability.

\clearpage
\bibliographystyle{IEEEtran}
\bibliography{mybib}

\end{document}